\documentclass [journal]{IEEEtran}
\usepackage{color,amsmath}
\usepackage {cite}

\usepackage{amssymb}
\usepackage{mathdots}
\usepackage[capitalise]{cleveref}
\usepackage{soul}
\usepackage[pdftex]{graphicx}
\graphicspath{{./figs/}{./picture/}}

\DeclareGraphicsExtensions{.pdf,.jpeg,.png}
\usepackage[caption=false,font=footnotesize]{subfig}

\begin{document}

\title{Realization of Reconfigurable Intelligent Surfaces with Space-Time Coded Metasurfaces}

\author{Mehdi Gholami, Soheil Khajavi, Mohammad Neshat, \IEEEmembership{Member, IEEE}, Simon Tewes, \IEEEmembership{Member, IEEE} and Aydin Sezgin, \IEEEmembership{Senior Member, IEEE}
\thanks{This work was supported in part by the German Federal Ministry of Education and
Research (BMBF) project 6G-ANNA [grant agreement number 16KISK095] and in part by the Deutsche Forschungsgemeinschaft (DFG, German Research Foundation) Project–ID287022738 TRR 196 (S03). \textit{(Corresponding author: Mohammad Neshat)}}
\thanks{Mehdi Gholami and Soheil Khajavi are with the  School of Electrical and Computer Engineering, College of Engineering, University of Tehran, Tehran, Iran}
\thanks{Mohammad Neshat is with the School of Electrical and Computer Engineering, College of Engineering, University of Tehran, Tehran, Iran, and also with the Centre for Wireless Innovation (CWI), Queen’s University Belfast, Belfast, UK (e-mail: m.neshat@qub.ac.uk)}
\thanks{Simon Tewes and Aydin Sezgin are with the Institute of Digital Communication Systems, Ruhr University Bochum, Bochum, Germany (e-mail: aydin.sezgin@rub.de)}}

\maketitle
\begin{abstract}
This paper presents experimental realization of a reconfigurable intelligent surface (RIS) using space-time coded metasurfaces to enable concurrent beam steering and data modulation. The proposed approach harnesses the capabilities of metasurfaces, allowing precise temporal control over individual unit cells of the RIS. We show that by employing proper binary codes manipulating the state of unit cells, the RIS can act as a digital data modulator with beam steering capability. We describe the experimental setup and computational tools, followed by validation through harmonic generation and investigation of beam steering and data modulation. Additionally, four digital modulation schemes are evaluated. By implementing customized binary codes, constellations under varying conditions are compared, showcasing the potential for real-world applications. This study offers new insights into the practical implementation of RIS for advanced wireless communication systems.
\end{abstract}


\section{Introduction}
The 21st century has witnessed a remarkable increase in the demand for telecommunication services, establishing it as one of the most significant expansions in recent history. The latest advancement in mobile communication, known as the fifth generation (5G), has addressed this issue by implementing large-scale multiple-input multiple-output (MIMO) antennas that utilize fully digital modulation \cite{agiwal2016next}. Recently, 5G has expanded its frequency range to include the mm-wave spectrum \cite{dehos2014millimeter}. But as time passes, the need for connecting people worldwide with improved speed, latency, and coverage is increasing as new technologies emerge. Applications such as virtual reality, autonomous driving, and tactile internet require users to be linked to networks at an ideal level, which delivers the wireless communication technology to the next generation, 6G \cite{tariq2020speculative}.

There are various methods to improve the service in the future generation. One approach is to elevate the frequency of the carrier wave to the terahertz range \cite{harter2020generalized, hosseininejad2018mac, hosseininejad2018reconfigurable}. Also, there are researches on the use of extremely large aperture and ultra massive MIMO (xMaMIMO, UMMIMO) antennas in this domain \cite{han2018ultra, amiri2018extremely}. There are also many researches on using optically controlled reconfigurable switches \cite{li2024,patron2014}.
These technologies primarily emphasize on the receivers and transmitters, and in some instances require sophisticated and expensive equipment. \\
Another innovative advancement is the emergence of reconfigurable intelligent surfaces (RISs) \cite{liu2021reconfigurable,elmossallamy2020reconfigurable,basar2019wireless,wu2019towards,alamzadeh2023,el2023,2aydin2023IEEE,alamzadeh2022,aydin2023IEEE}, which has drawn much attention and has the potential to be a paradigm shift in this field of research.
Contrary to other methods, RIS is a cost and energy efficient semi-passive component that aims to optimize the channel and has the ability of modulation and controlling the direction of the wave in a wide range of spectrum, from microwave to terahertz and visible light \cite{yang2022terahertz, aboagye2022ris}.

RIS is generally made of an array of programmable unit cells on a metasurface \cite{cui2014coding,zhang2018space}. By manipulating the phase and amplitude of the reflected wave from each cell, it is feasible to modify the wavefront of the total reflected wave, making it a highly practical tool in wireless communication.
The traditional RIS-assisted wireless network solutions primarily employed phase shifters to control the electromagnetic wave at each cell \cite{yang2020coverage,yu2019miso,di2020hybrid,huang2019reconfigurable}. Despite its effectiveness, phase shifters necessitate a high level of complexity and cost. Eventually, a different technique so called space-time coding (STC) was introduced as a replacement for the phase shifters. The application of space-time coding on a surface was initially introduced by Hadad et al. \cite{hadad2016breaking,hadad2015space}, in which they achieved space-time modulation of the electromagnetic wave. Following that, investigations were conducted to achieve the manipulation of electromagnetic waves by time-domain digital coding metasurfaces \cite{dai2018independent,zhang2018space,Liu21,Pang22}. This technique involves designating two states for each cell in the RIS. By applying a binary code to each cell and switching between the states, it becomes feasible to manipulate the amplitude and phase of each cell, so achieving precise control over the beam shape.

\begin{figure}[t]
\centering
\includegraphics[width=3.5in]{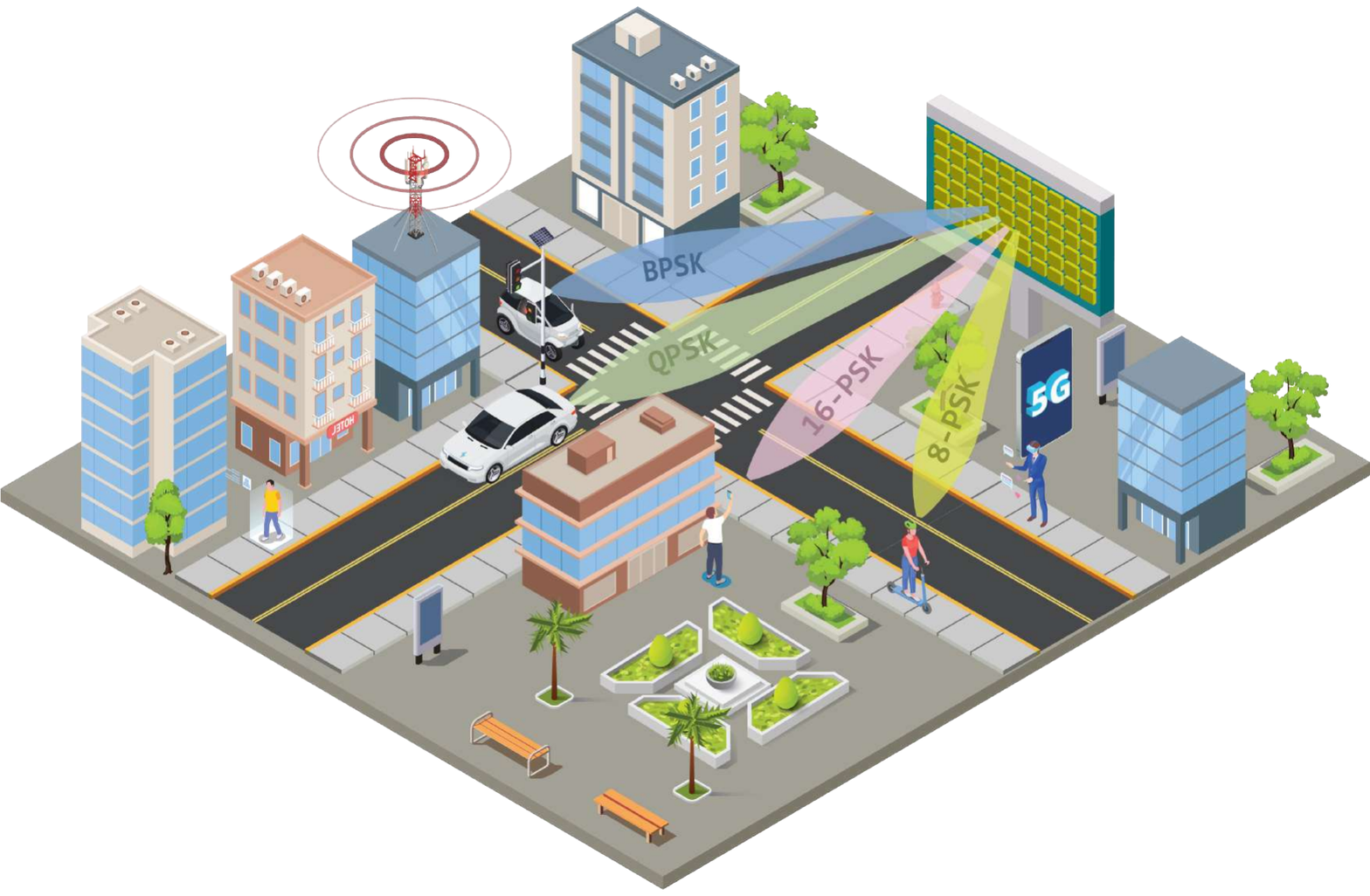}
\caption{Illustration of a RIS as an access point within a cell-free network of a smart city. The RIS, mounted on a billboard, modulates and steers beams tailored to individual users. 
} 
\label{senario}
\end{figure}

In this paper, our aim is to experimentally demonstrate the integration of various modulation schemes with beam steering capability via RIS, based on the principles of space-time coding technique. Our methodology employs a RIS composed of discrete unit cells. These cells are manipulated via electronic switches to alter their operational state, facilitating simultanious digital modulation and beamforming. 
Fig~\ref{senario} illustrates a practical use case of our proposed concept for implementation of different modulation schemes at specific directions within a smart city and delivering service to users located in areas where the antenna signal is obstructed.

The subsequent sections of the article are structured as follows. Section II provides an overview of our methodology, encompassing the vector representation of bits, the introduction of the utilized RIS, the configuration of measurements, and the computational tools employed throughout the experiments. In Section III, we present the outcomes of the measurements, commencing with harmonic generations. Subsequently, we detail the results of beam steering, offering comparisons to simulated outcomes. Following this, we present the calculated constellations for four distinct modulation schemes achieved through the applied codes. Finally, we showcase the radiation pattern and constellation of a QPSK modulation in various angular directions, followed by concluding remarks in Section IV.

\section{Methodology}
In this section, we begin by revisiting our previous investigation into the vector representation of bits. Following that, we detail the utilized RIS and the configuration setup employed for conducting the experiments. Subsequently, we introduce a series of software programs we developed for each phase of the experiments.
\begin{figure}[b]
\centering
\subfloat[]{\includegraphics[width=1.7in,page=1]{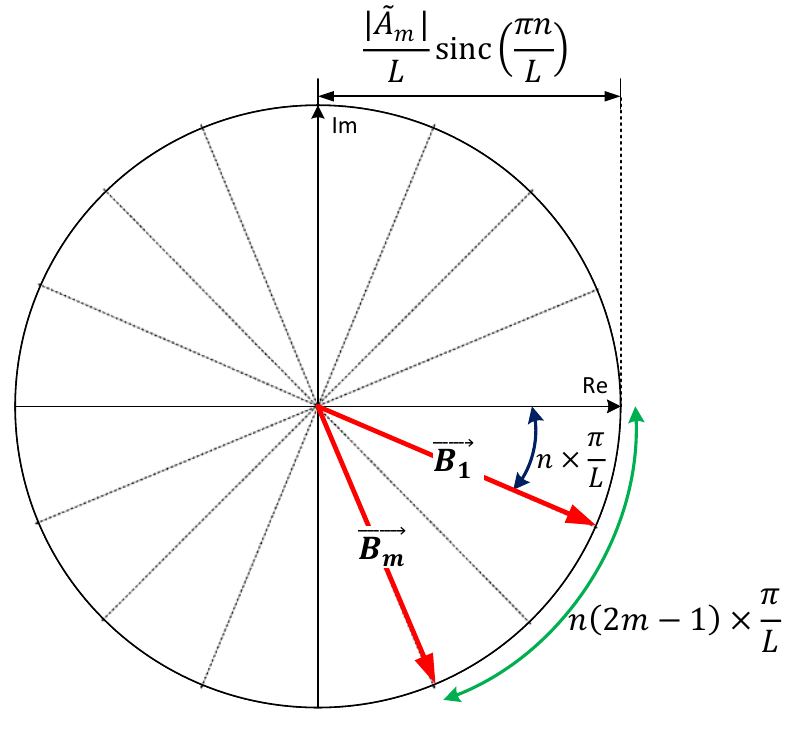} \label{vector}}
\subfloat[]{\includegraphics[width=1.7in,page=1]{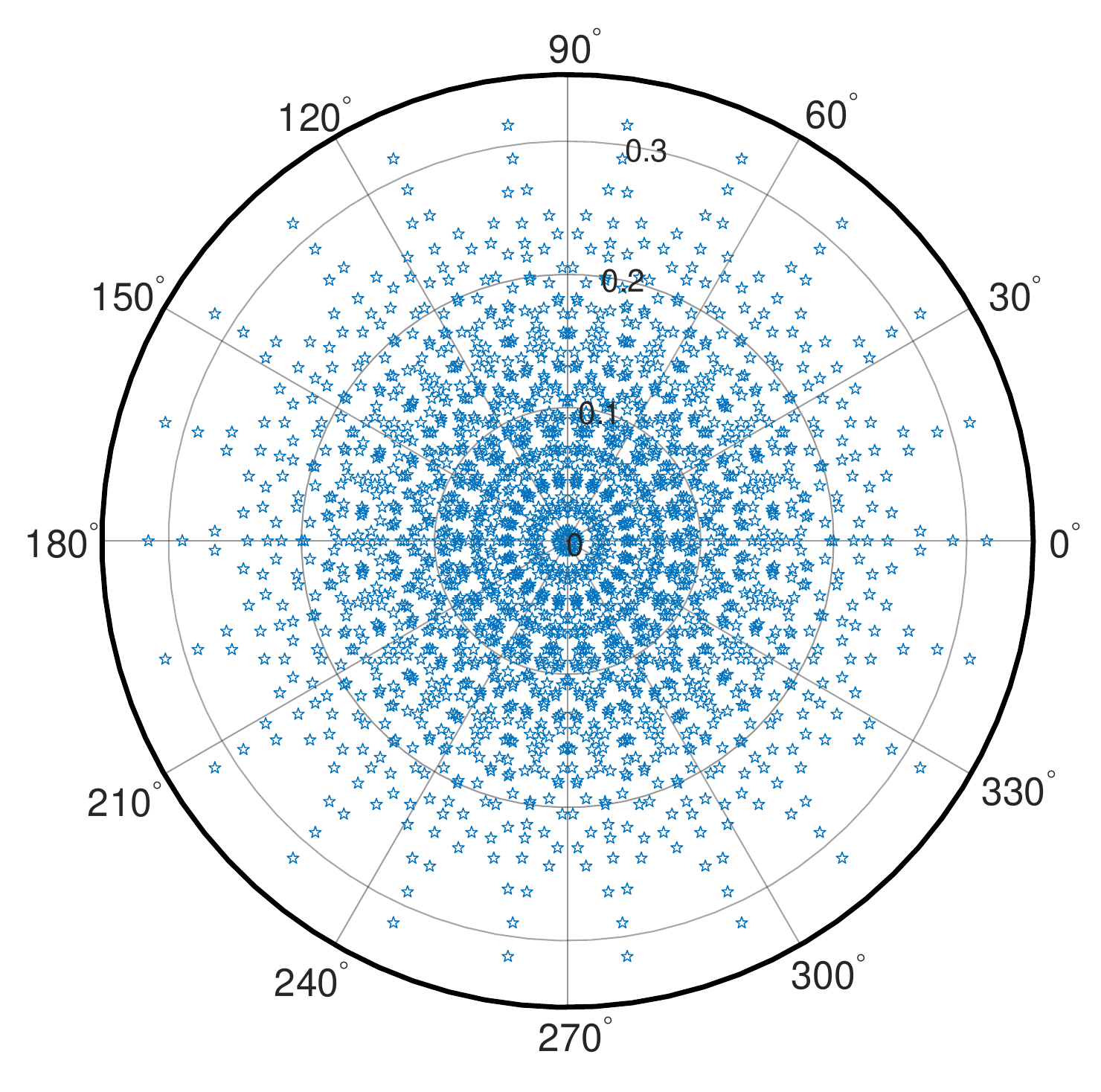} \label{constellation}}
\caption{ 
\protect\subref{vector} Vector representation of bits for $n^{th}$ harmonic in time coding. \protect\subref{constellation} Amplitude-phase pattern for the first harmonic generated by all possible 11-bit codes.}
\label{vector_constellation}
\end{figure}
\subsection{Vectors Representation of Control Bits}
Shaping the wavefront of an electromagnetic wave holds significant importance within communications. This task necessitates precise control over both amplitude and phase of the wave. Multiple methods exist for the manipulation of amplitude and phase, either at the source or through the utilization of reflective devices. A simple approach to achieve this objective is via space-time coding that involves temporal alternation of unit cells, each encoded with specific code sequences, across their distribution area. By modulating the signal with binary codes, akin to on-off switches, it becomes evident that this technique can modify the signal's amplitude and phase in a desired fashion. Here, we elucidate our prior discovery of representing codes via vectors \cite{gholami2022direct}, consequently enhancing the selection process for the appropriate codes aligned with our objectives.

When a binary code with a length of $L$ bits, recurring periodically with a period of $T_0$, is multiplied by a sinusoidal signal, it induces the generation of distinct harmonics. The manipulation of these harmonics' amplitude and phase is achievable through the selection of various codes and their corresponding bit lengths. $m^{th}$ bit within the $L-$bit code can be associated with a vector $\Vec{B}_m$ in the complex plane as shown in Fig. \ref{vector}. The vector $\Vec{B}_m$ at harmonic order $n$ exhibits the amplitude and phase as \cite{gholami2022direct}

\begin{subequations}
\label{eq.Bm}
\begin{align}
  \label{eq.Bm.amp}
  \left\vert{}\Vec{B}_m\right\vert{}&=\frac{\left\vert{}\tilde{A}_m\right\vert{}}{L}\textrm{sinc}\left(\frac{\pi{}n}{L}\right)
 \\
  \label{eq.Bm.pha}
  \angle \Vec{B}_m &=-n\left(2m-1\right)\times \frac{\pi}{L}.
\end{align}
\end{subequations}

\noindent where $\tilde{A}_m$ is a complex number determining the state of the $m^{th}$ bit, e.g. $\tilde{A}_m=1\angle0, 1\angle\pi ~\textrm{or}~ 0$. The complex amplitude of the $n^{th}$ harmonic is the summation of all associated vectors with the $L$ bits. Furthermore, for a specified code length, $L$, it is possible to graphically represent the resulting $2^L$ amplitude-phase points for any harmonic frequency using a polar plot. Fig.~\ref{constellation} illustrates amplitude-phase pattern for the first harmonic generated by 11-bit codes. Visual inspection of Fig. \ref{constellation} reveals the substantial degrees of freedom in both phase and amplitude for the first harmonic through time coding. In the subsequent section, we will elaborate on the employed metasurface to implement our desired codes across the cells via switching.

\subsection{Reconfigurable Intelligent Surface}
\begin{figure}[b]
\centering
\subfloat[]{\includegraphics[width=1.6in,page=1]{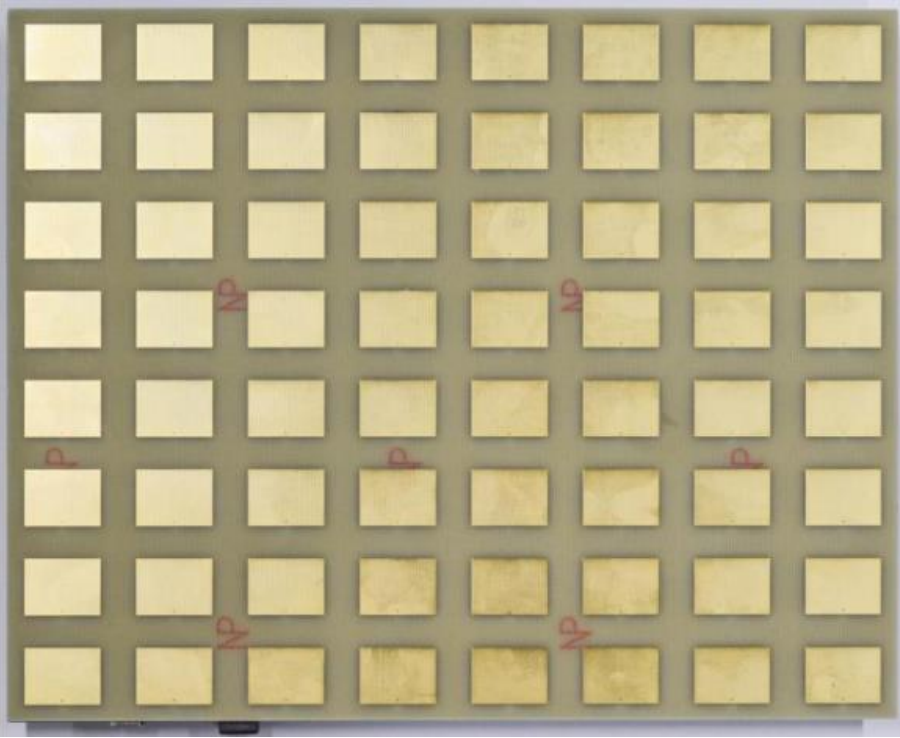} \label{ris.front}}
\subfloat[]{\includegraphics[width=1.6in,page=1]{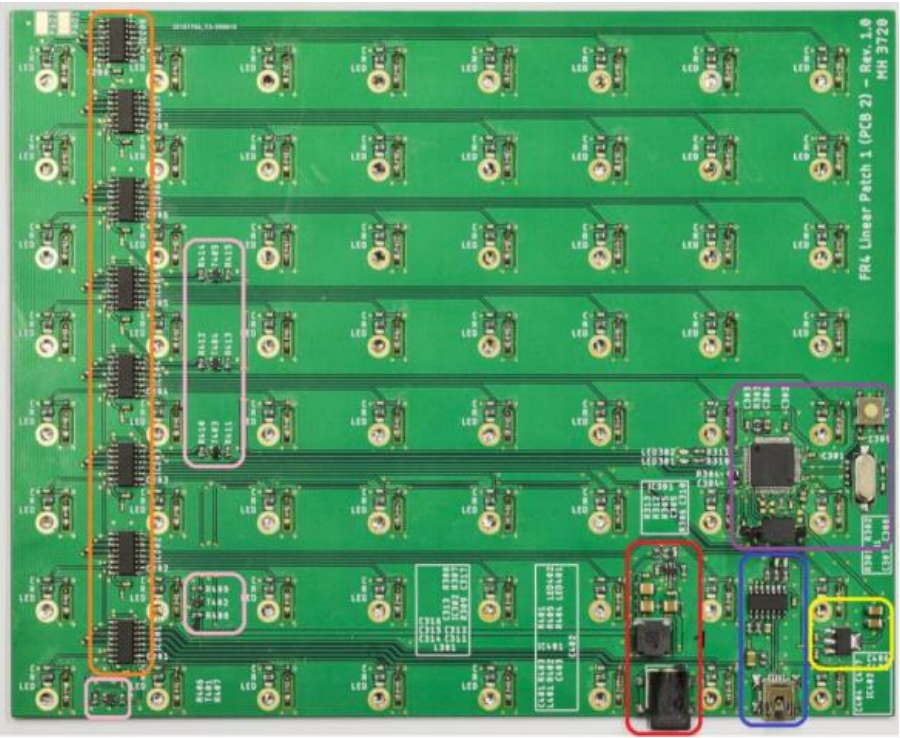} \label{ris.back}}
\caption{ 
\protect\subref{ris.front} Top view of the fabricated RIS PCB with 64 elements. \protect\subref{ris.back} Back view of the RIS PCB containing the circuitry needed to configure every cell. The circuits delineated by different colors are responsible for different tasks. Orange: Shift registers that facilitate the conversion of diode biases from serial bits to parallel lines, Pink: Voltage converter; transforms the voltage to the bias level, Red: 5 V DC/DC converter, Blue: USB to UART convertor, Yellow: Linear regulator 3.3 V, Purple: Microcontroller with a reset button and serial ports for programming.} 
\label{ris}
\end{figure}

The central component in our investigation on space-time coding is a RIS. We developed a metasurface by adapting the unit cells proposed in \cite{kaina}. Fig. \ref{ris} shows the fabricated RIS that is composed of 64 unit cells arranged in an $8\times8$ array, forming overall dimensions of $160\times200$  mm. To configure each unit cell, a microcontroller placed on the back side utilizes a serial connection for programming purposes. The phase of the reflection coefficient of each unit cell can be controlled by a binary bit upon illumination by linearly polarized waves. Every cell consists of a reflecting patch antenna backed by the ground plane. The length of the patch determines the main resonance frequency, which corresponds to half the wavelength. Through the utilization of a PIN diode, the effective length of the patch becomes variable, thereby inducing a modification in the reflection coefficient at the desired frequency. The diode, situated behind the patch, makes a connection between the patch and the ground plane through a via. Such via is displaced from the center of the patch by a minor offset. The resonance frequency shifts to a higher value when the switch is on, by an incremental in the via offset. 

\subsection{Measurement Setup}
\begin{figure}[t]
\centering
\includegraphics[width=3.5in]{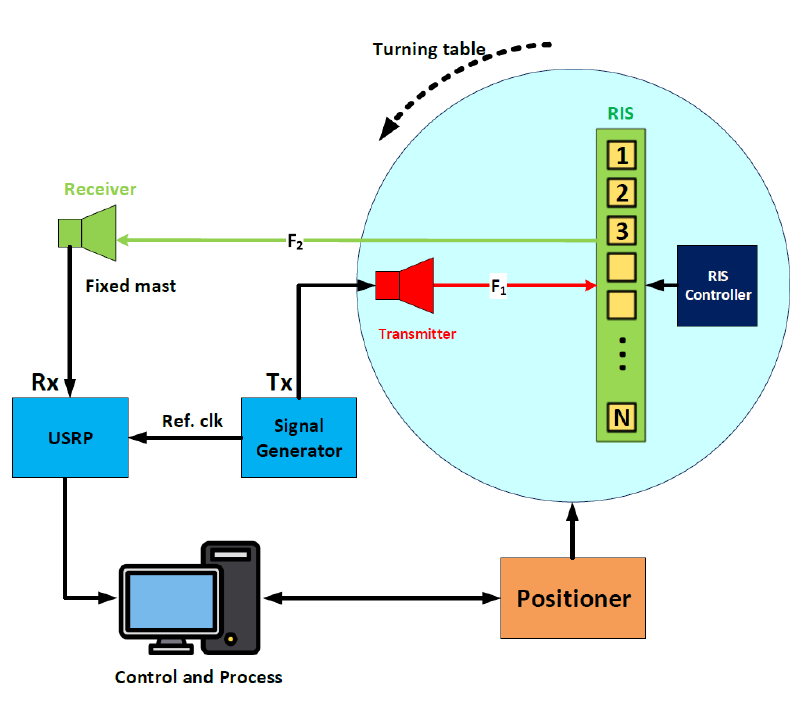}
\caption{Block diagram of the radio link measurement setup using a space-time coded RIS.
} 
\label{setup}
\end{figure}

\begin{figure}[b]
\centering
\includegraphics[width=3.5in]{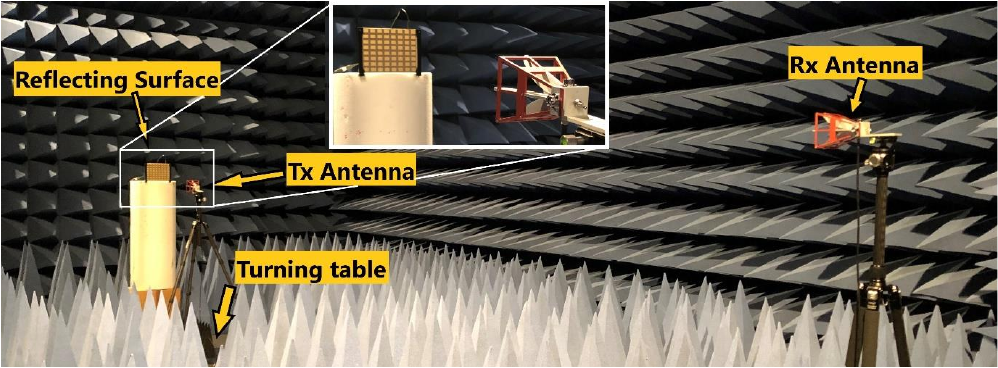}
\caption{Configuration of the radio link measurement setup in an anechoic chamber.
} 
\label{experiment}
\end{figure}

Fig.~\ref{setup} depicts the block diagram of the radio link measurement setup using a space-time coded RIS. The actual experimental arrangement within the anechoic chamber is shown in Fig.~\ref{experiment}. In this setup, the RIS is placed at the center of a rotary table. A microcontroller transmits data bits to the RIS and, via a LAN cable, allows a computer to program the controller remotely from outside the anechoic chamber. The computer executes a code that defines the modulation scheme, beam angle, and data for transmission. A Tx horn antenna with vertical polarization is positioned on a rotary table, aligned with the axis of the RIS with a slightly offset downward to prevent the scattered wave blockage. The position of the illuminating antenna remains fixed with respect to the RIS throughout all measurements. The Tx antenna is fed by a single tone via a signal generator operating at 5.3 GHz. The output clock from the signal generator is connected to a USRP radio receiver. An Rx horn antenna, attached to the radio receiver, captures the incoming wave. Following downconversion, the signal is then conveyed to the external computer through a cable. The computer also controls the angle of the rotary table, stores received data samples, and labels them with the corresponding angular position of the rotary table. 

\subsection{Software Tool}
To facilitate the measurements, three sets of software programs were developed and employed across all experiments. \\
\begin{enumerate}
  \item Binary code generation:  This program generates binary codes to control each cell at every time step. Any bit shifting and number of code repetition is determined in this program. The generated codes are stored in the on-board memory and utilized upon the activation of the RIS.

  \item Test environment control: This program generates the activation command for the RIS, configures the parameters of the signal generator (such as power and central frequency), and coordinates commands for the rotation of the table to achieve the desired angles. 
  
  \item Signal processing: This program utilizes the stored signals from the "Test environment control" program, and through subsequent processing, produces two outputs:
  \begin{enumerate}
     \item Received signal spectrum by applying Fourier transform on the received signals.
     \item Constellation diagram of the codes by filtering the desired harmonics and sampling the symbols.
     \end{enumerate}
\end{enumerate}

\section{Measurement Results}
In this section, we report the experimental results of the proposed wireless link based on space-time coded RIS. Initially, we verify the generation of harmonics via space-time coding, and demonstrate beam steering. Following that, we illustrate the application of this technique for modulating electromagnetic waves. Finally, we establish the feasibility of beam steering and data modulation simultaneously.

\begin{figure}[h]
\centering
\subfloat[]{\includegraphics[width=3.2in,page=1]{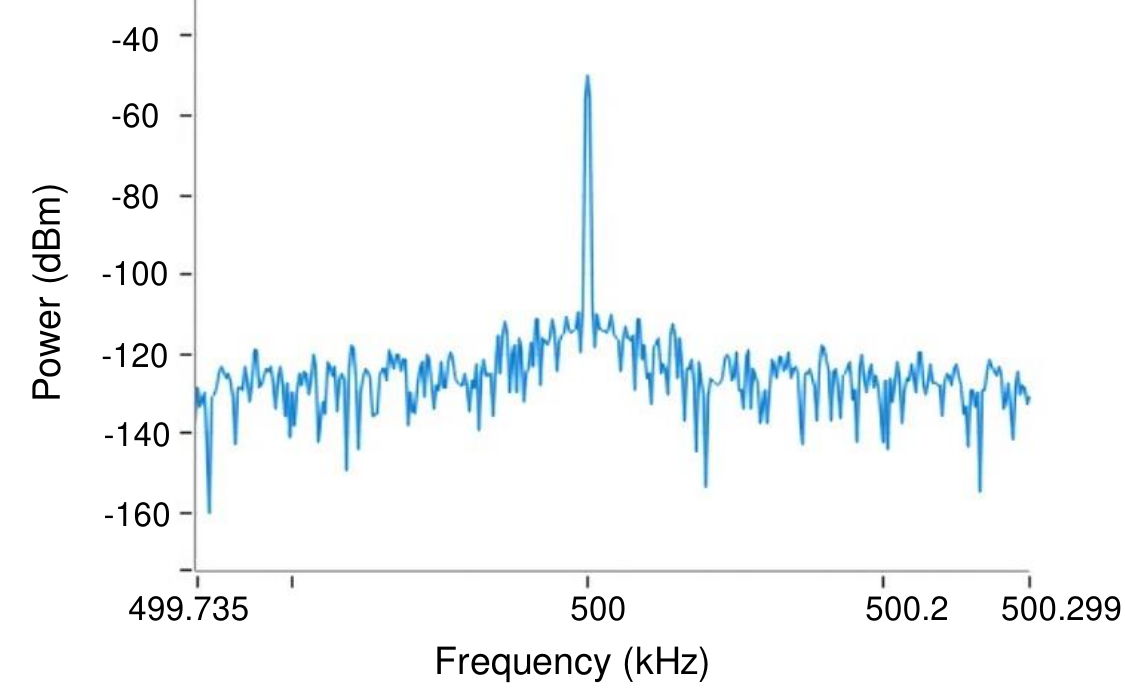} \label{main_harmonic}} 
\\
\centering
\subfloat[]{\includegraphics[width=3in,page=1]{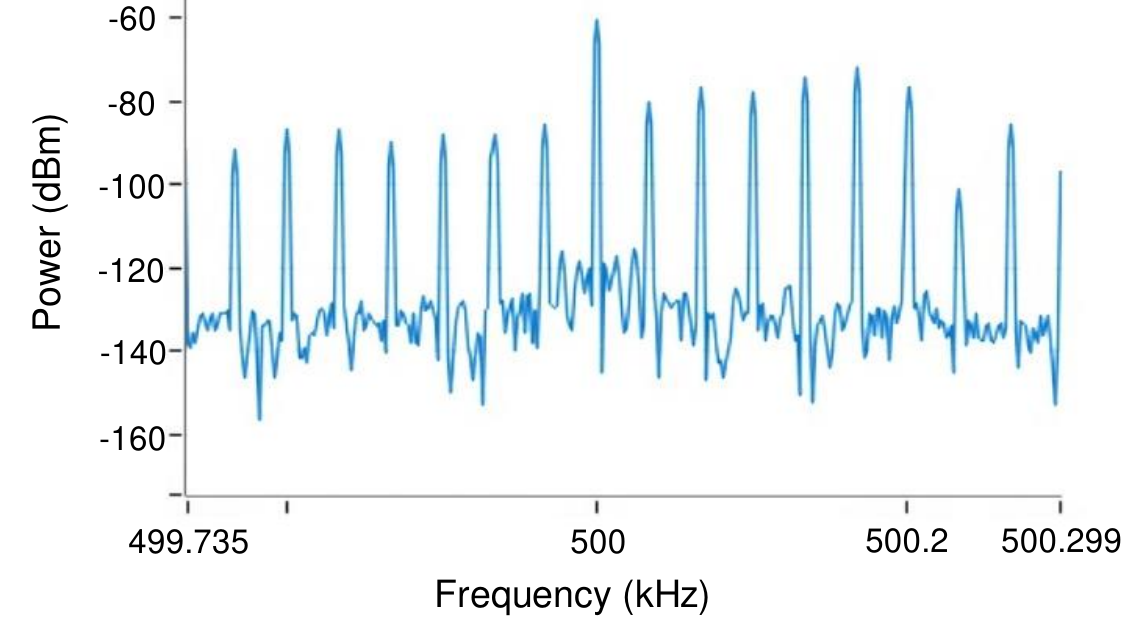} \label{generated_harmonics}}
\caption{ 
Spectrum of the received signal when \protect\subref{main_harmonic} all cells are at off state,  \protect\subref{generated_harmonics} the binary code 0000 0001 is periodically applied to the cells of the RIS. } 
\label{harmonics}
\end{figure}

\subsection{Generation of Harmonics due to Time Coding}
We first illuminate the RIS with a single tone while all the cells are at off state. Fig.~\ref{main_harmonic} shows the spectrum of the received signal at the base band. At the base band, the zero harmonic (single tone carrier) is detected at zero frequency or DC. In order to overcome the noise and distortion of the downconvertor at DC, 500 kHz offset is added. Therefore, as seen in Fig. \ref{main_harmonic}, the transmitted single tone is detected at 500 kHz in the base band instead of at DC, and the RIS operates as a simple reflector.

In the next step, we apply an identical binary code with 8 bits, 0000 0001, to all cells simultaneously. The switching time duration for each bit is $\tau=3.74$ ms. Given the code length of $L=8$ bits, the time period of the code is $T=29.92$ ms. Therefore, the frequency of the first harmonic is at $f_1=1/T=33.42$  Hz. Such low harmonic frequency is a consequence of the microcontroller limitation, which transmits data serially. Parallel data transmission can result in shorter switching time and  higher harmonic frequency. According to the vector representation of Fig. \ref{vector}, one expects to have many harmonics for the chosen code as is evident in the spectrum of the received signal in Fig. \ref{generated_harmonics}. As predicted above, the harmonic distance is close to 33.42 Hz.

\subsection{Beam Steering}
In this section, our objective is to achieve beam steering of harmonics through meticulous control of the codes applied to the cells. Referring to the vector representation depicted in Fig. \ref{vector}, it becomes evident that shifting the order of bits in a code results in a corresponding change in the phase of the harmonic. The phase change for $n^{th}$ harmonic, resulting from an $s$-bit shift, can be quantified as

\begin{equation}
    \Delta\phi_{n,s}=\frac{2\pi n s}{L}
    \label{phase_change}
\end{equation}

\noindent where $s=1, ..., L$. It is easy to show that the steering angle for the $n^{th}$ harmonic following an $s$-bit shift consecutively between   adjacent cells in a RIS with the distance of half a wavelength can be obtained as

\begin{equation}
    \theta_{n,s}=\arcsin\left(\frac{2ns}{L}\right)
    \label{steering_angle}
\end{equation}

\begin{figure}[b]
\centering
\includegraphics[width=3.5in]{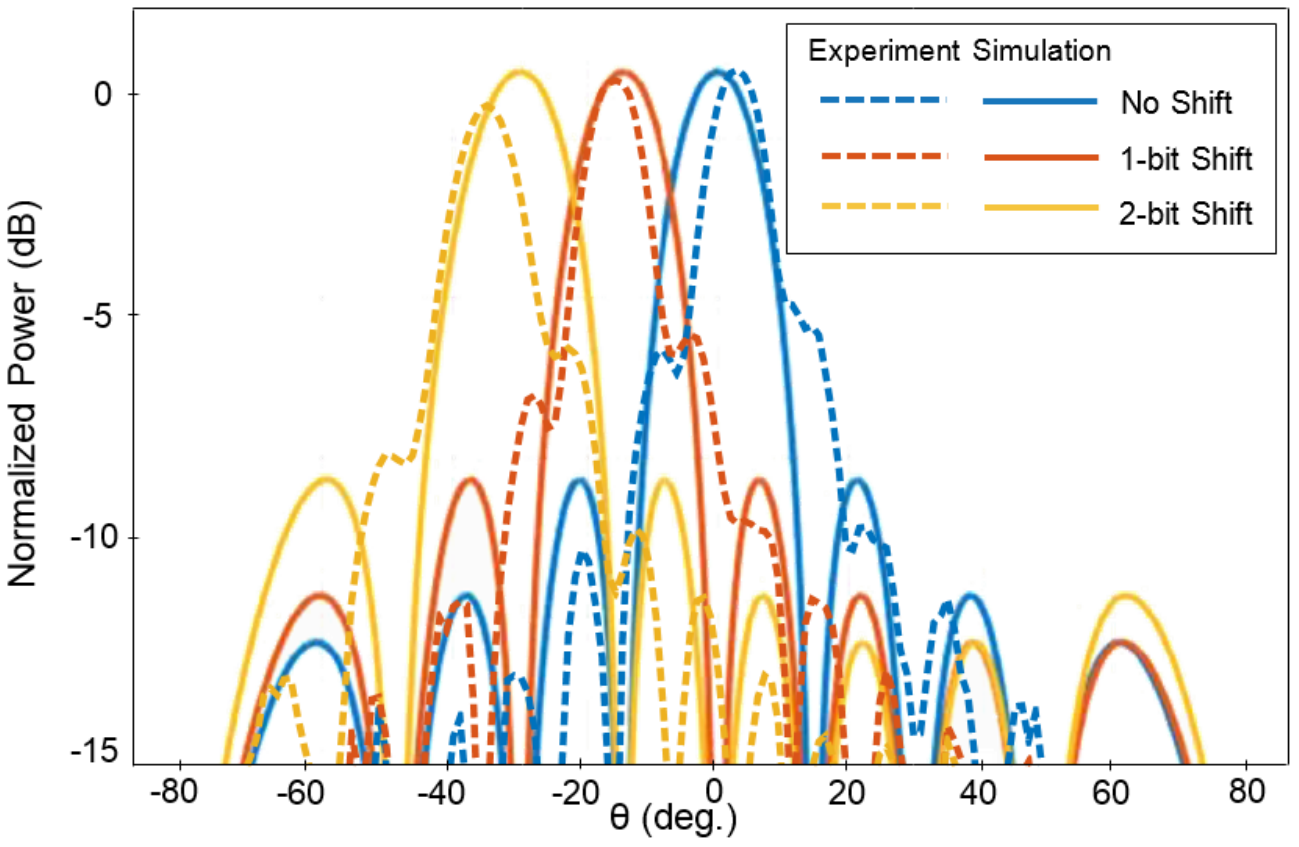}
\caption{Simulation of array factor and measurement  of reflected pattern from the RIS for the first harmonic. Blue: same code (with no shift) is applied to all columns, red: 1-bit shift, yellow: 2-bit shift between adjacent columns.
} 
\label{Steering}
\end{figure}

To demonstrate the capability of beam steering, we first apply a specific code to all the cells and observe the reflected pattern from RIS for the first harmonic. Subsequently, we introduce a 1-bit shift in 8-bit codes applied to each column. The choice of column wise shifts is made to achieve beam steering in one direction, simplifying the measurement process. Then, we apply a two-bit shift in the columns and observe the reflected pattern. Fig.~\ref{Steering} compares the simulated array factor of such a scenario for a space-time coded linear cell array. It is evident from the solid-lines (simulation) that the steering angle for 1-bit and 2-bit shift column wise, is $\theta_{1,1}=14^\circ$ and $\theta_{1,2}=30^\circ$, respectively. Dashed lines in Fig.~\ref{Steering} show the measured reflected patterns from the space-time coded RIS and the measured steering angles are in consistent with the calculations and simulation. The observed decrease in peak power, correlating with the increment of the steering angle in measurements, can be attributed to the reduced effective area of the RIS.

\subsection{Data Modulation}
In the previous section, we established the efficacy of beam steering via space-time coding techniques in our experiments. Moving forward, our objective now is to illustrate the potential for data modulation through the application of space-time coding technique. This stage of the experiment unfolds across two different settings: initially, in an anechoic chamber, and subsequently, within a typical office space furnished with common equipment like computers, desks, seating, and more. Such varied environments enable us to demonstrate the flexibility of our methodology under practical conditions.
\begin{figure}[h]
\centering
\subfloat[]{\includegraphics[width=1.7in,page=1]{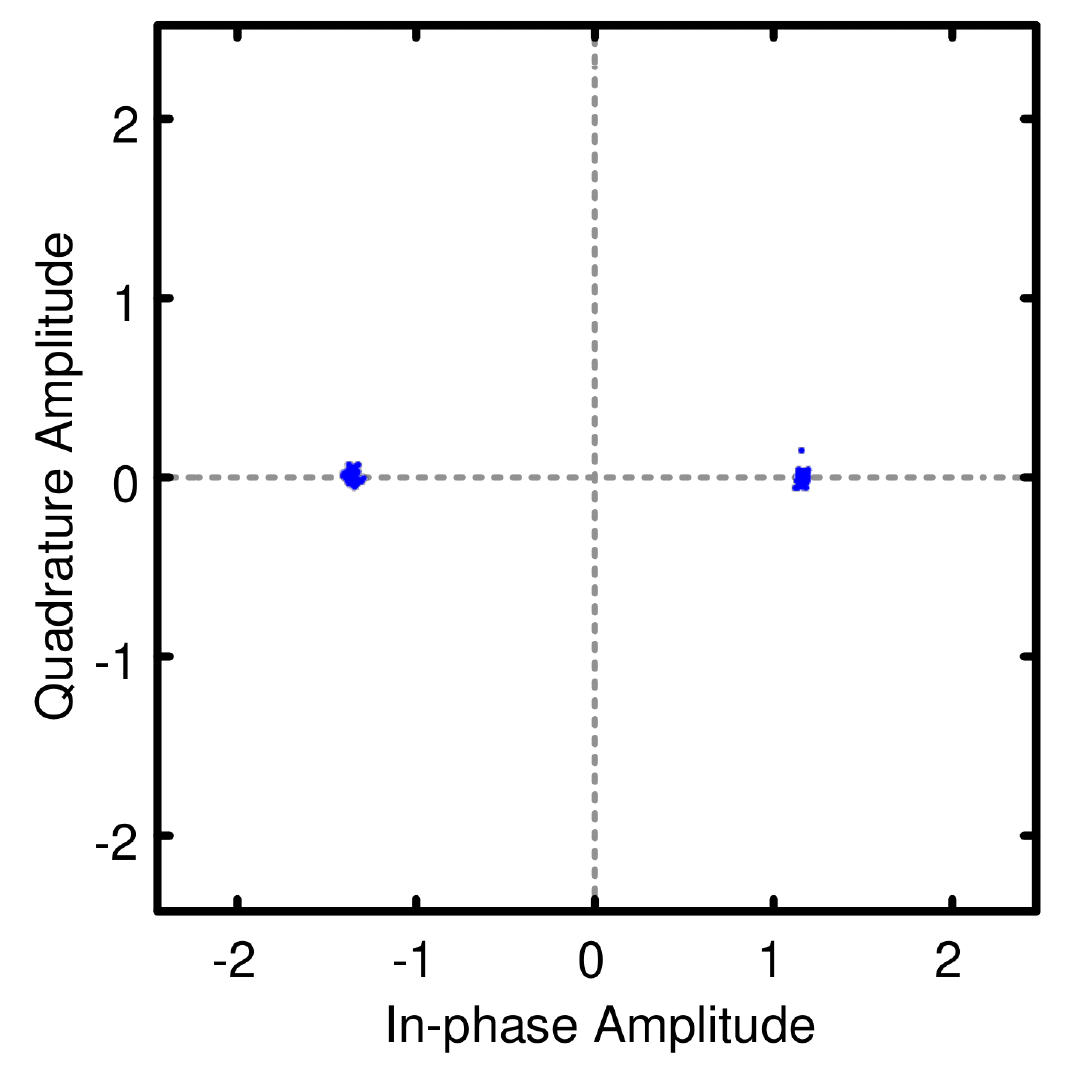} \label{BPSK2}}
\subfloat[]{\includegraphics[width=1.7in,page=1]{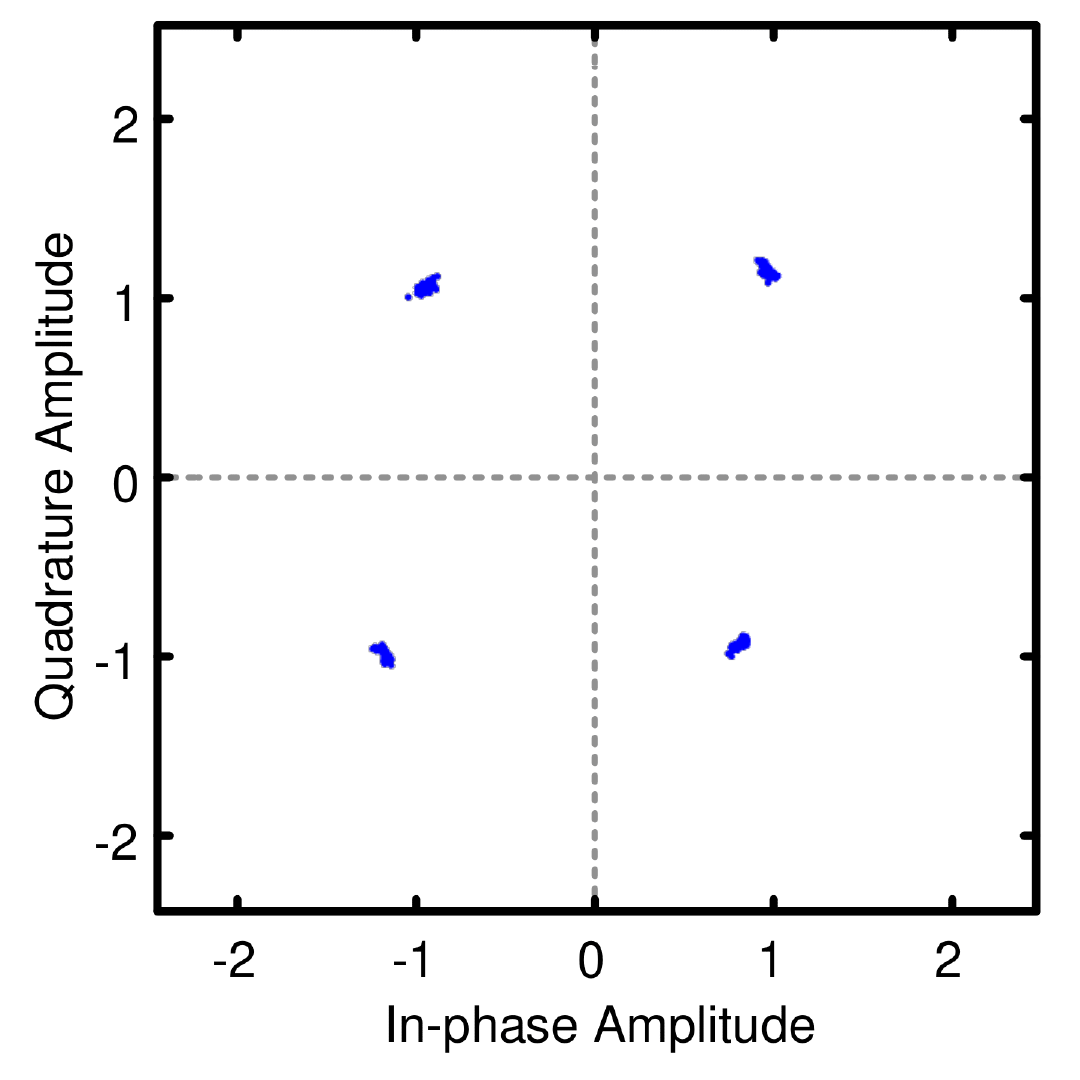} \label{QPSK2}}
\\
\subfloat[]{\includegraphics[width=1.7in,page=1]{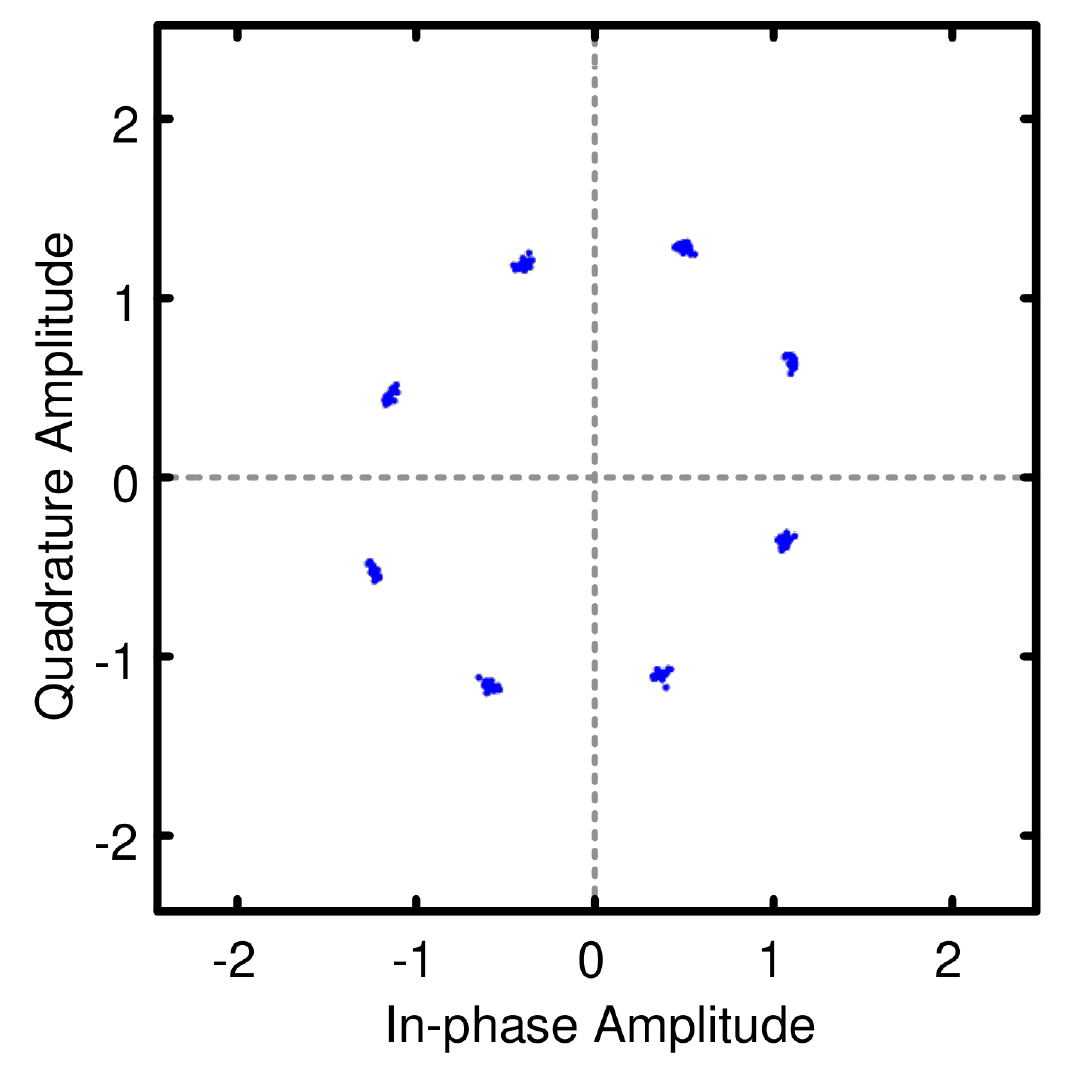} \label{8-PSK2}}
\subfloat[]{\includegraphics[width=1.7in,page=1]{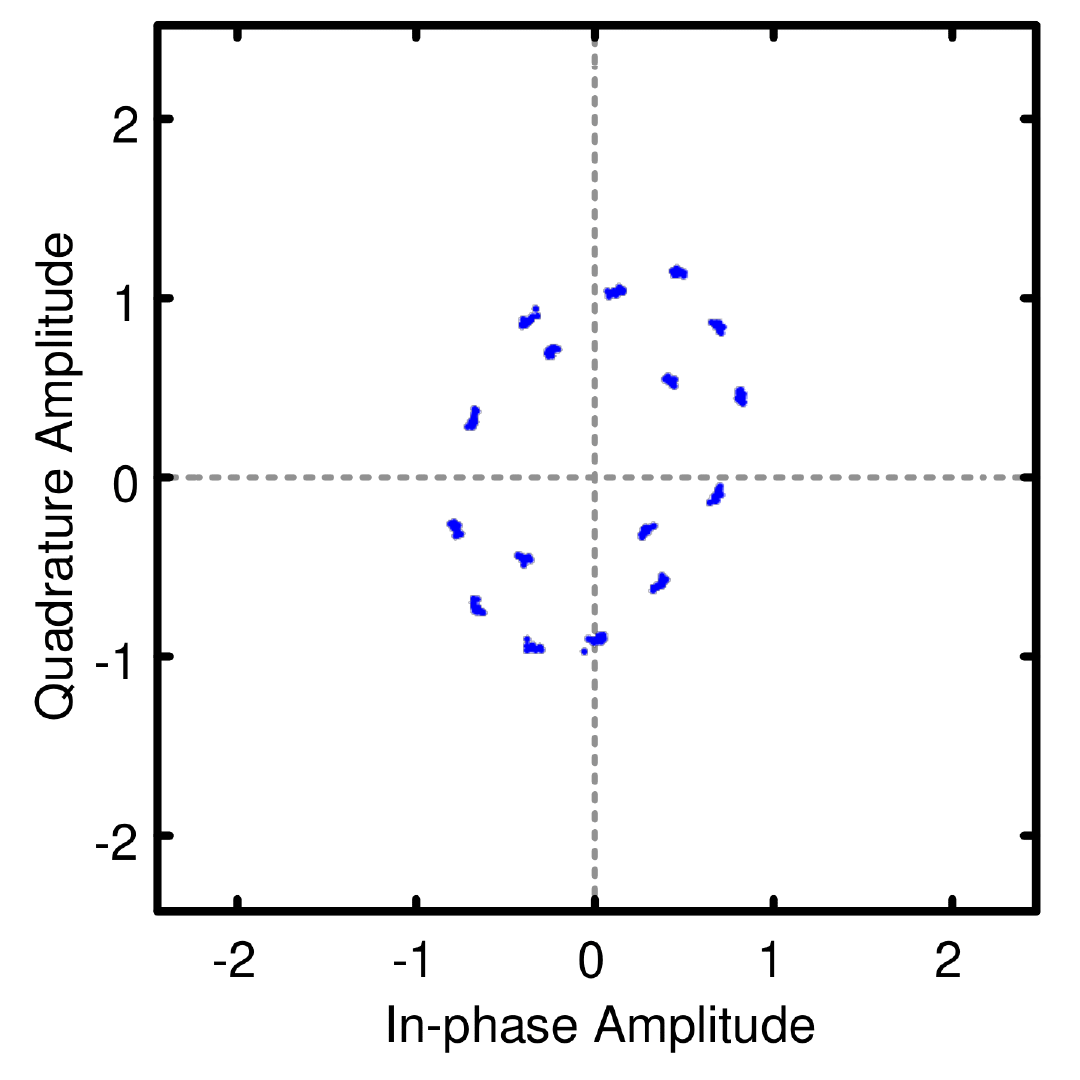} \label{16-PSK2}}

\caption{ 
Reconstructed constellations at the
receiver in the anechoic chamber for four modulation schemes: 
 \protect\subref{BPSK2} BPSK \protect\subref{QPSK2} QPSK \protect\subref{8-PSK2} 8-PSK \protect\subref{16-PSK2} 16-PSK.
} 
\label{Modulation_Antenna_Room}
\end{figure}
The experiment encompasses the implementation of four digital modulation schemes, BPSK, QPSK, 8-PSK, and 16-PSK within the first harmonic. Referring to the amplitude-phase pattern depicted in Fig. \ref{constellation}, and our previous research \cite{gholami2022direct} clearly indicates that multiple coding options are available for any given constellation. Our emphasis in this context is not on delving into digital communication intricacies, but rather on the generation of symbol constellations for the purpose of regenerating transmitted codes.

\begin{figure}[h]
\centering
\subfloat[]{\includegraphics[width=1.7in,page=1]{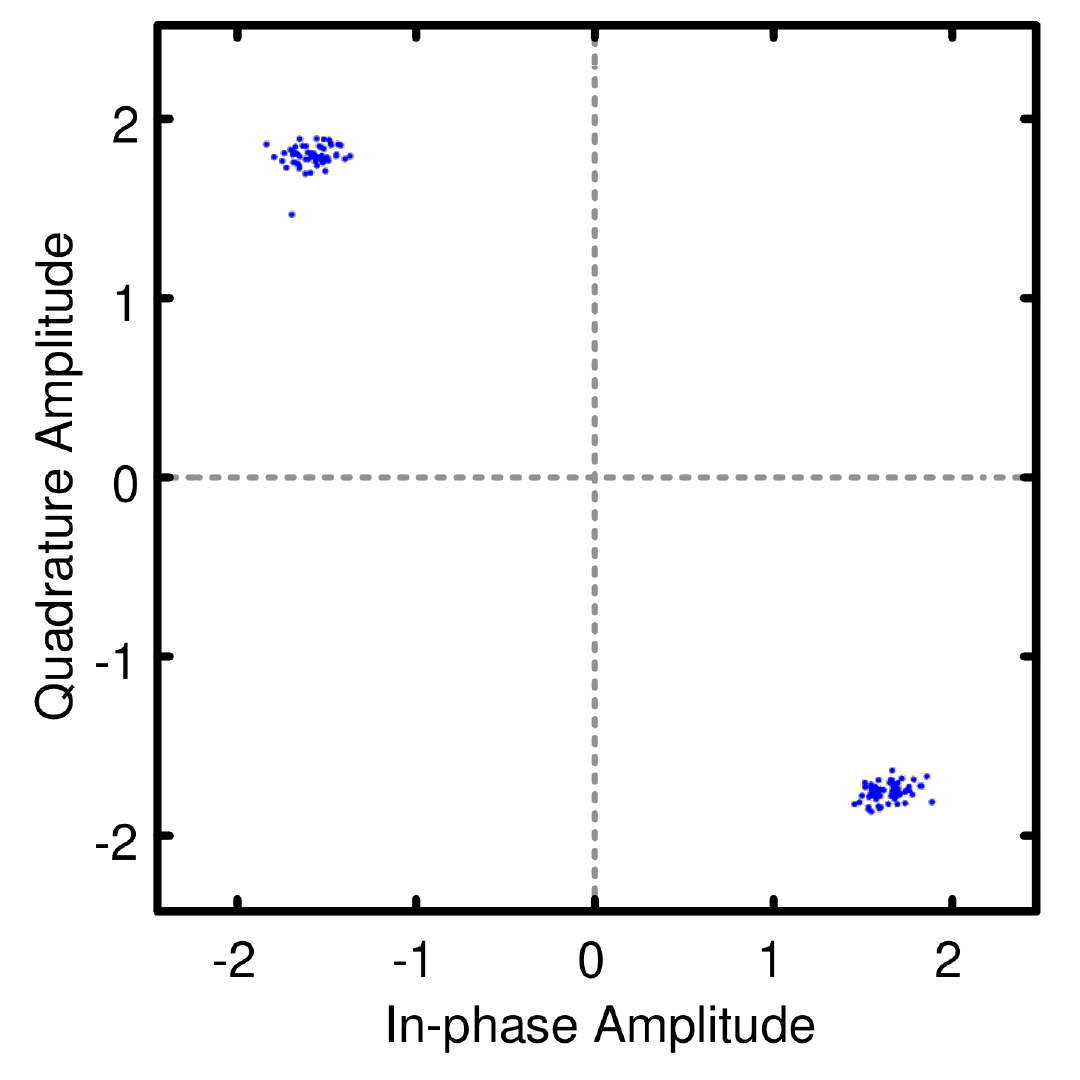} \label{BPSK1}}
\subfloat[]{\includegraphics[width=1.7in,page=1]{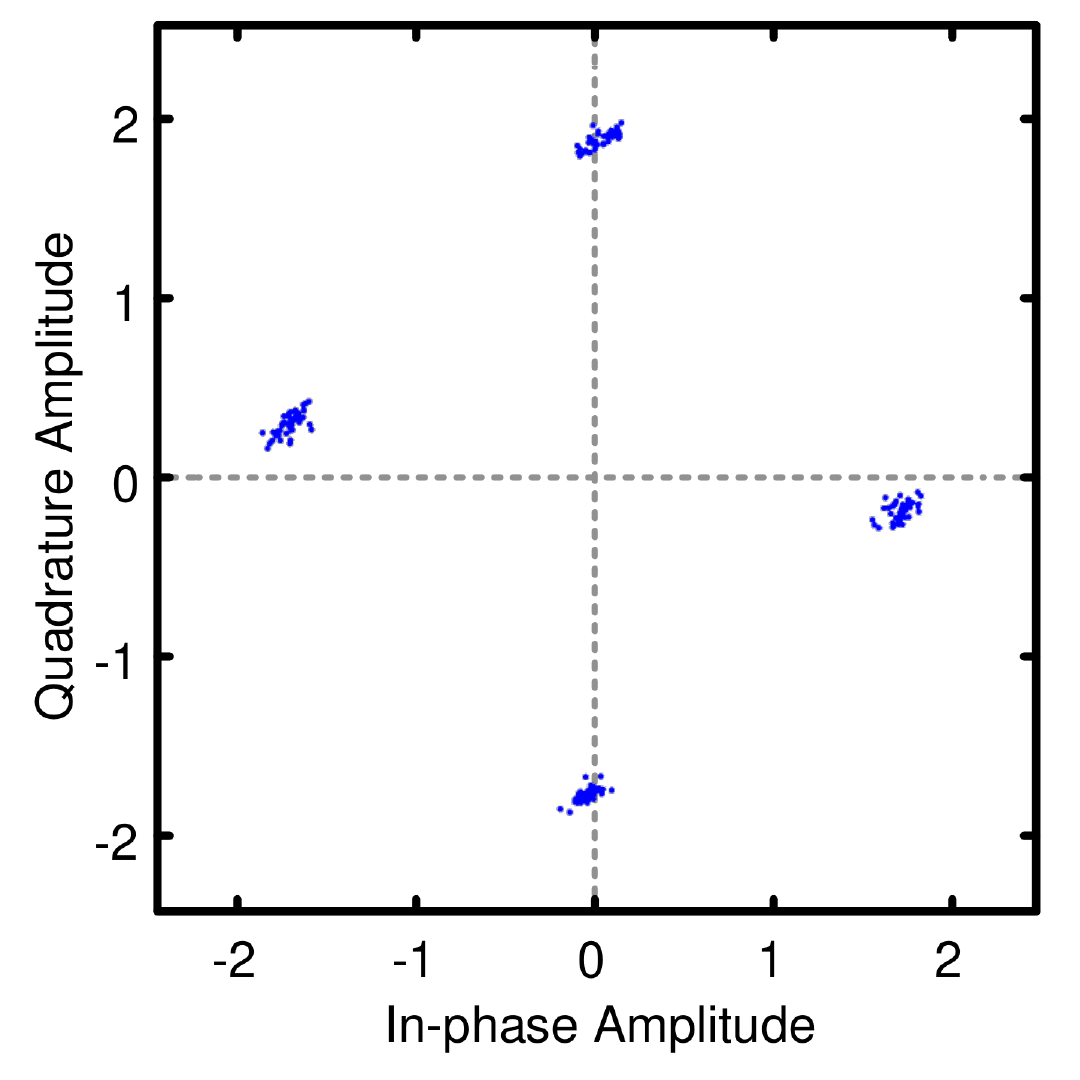} \label{QPSK1}}
\\
\subfloat[]{\includegraphics[width=1.7in,page=1]{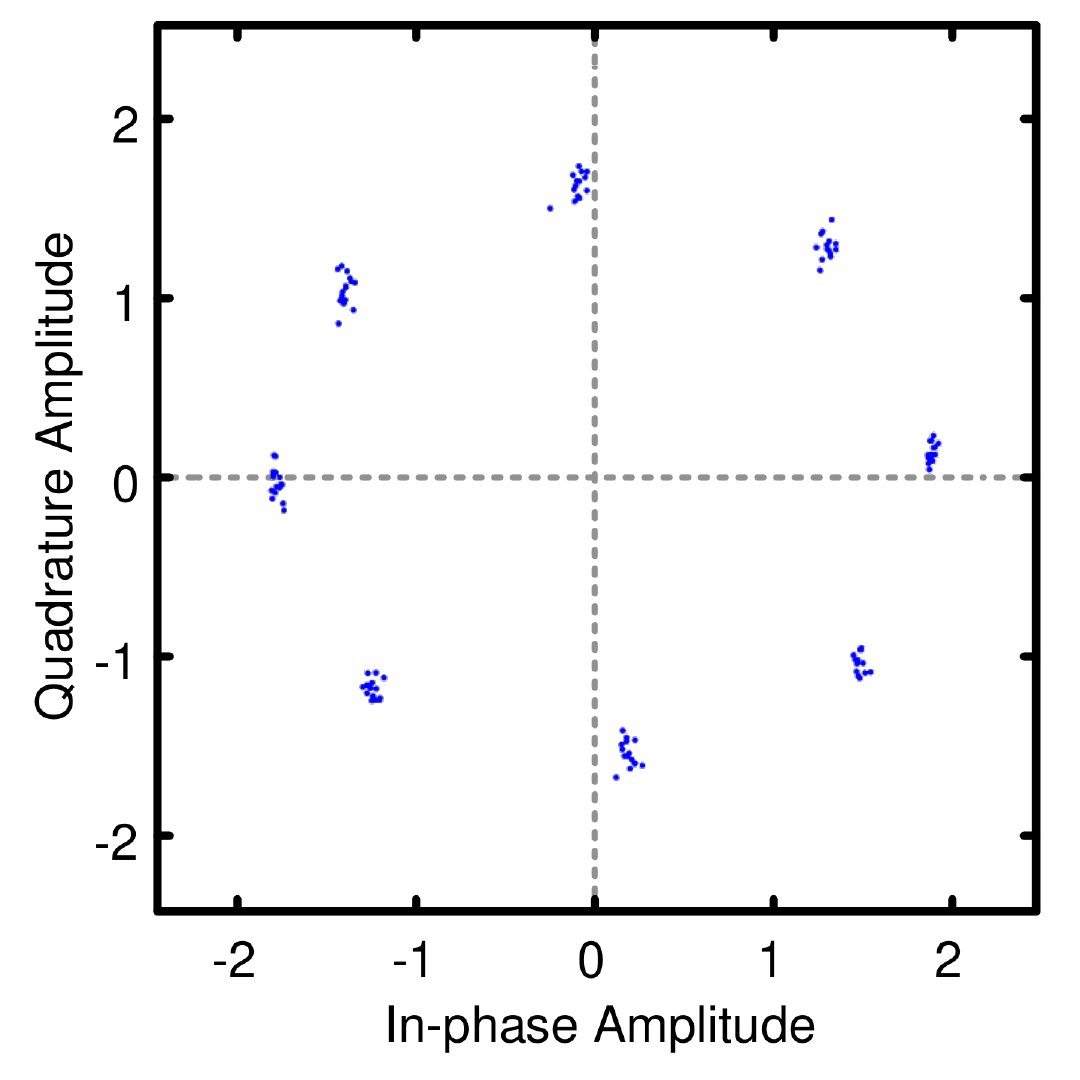} \label{8-PSK1}}
\subfloat[]{\includegraphics[width=1.7in,page=1]{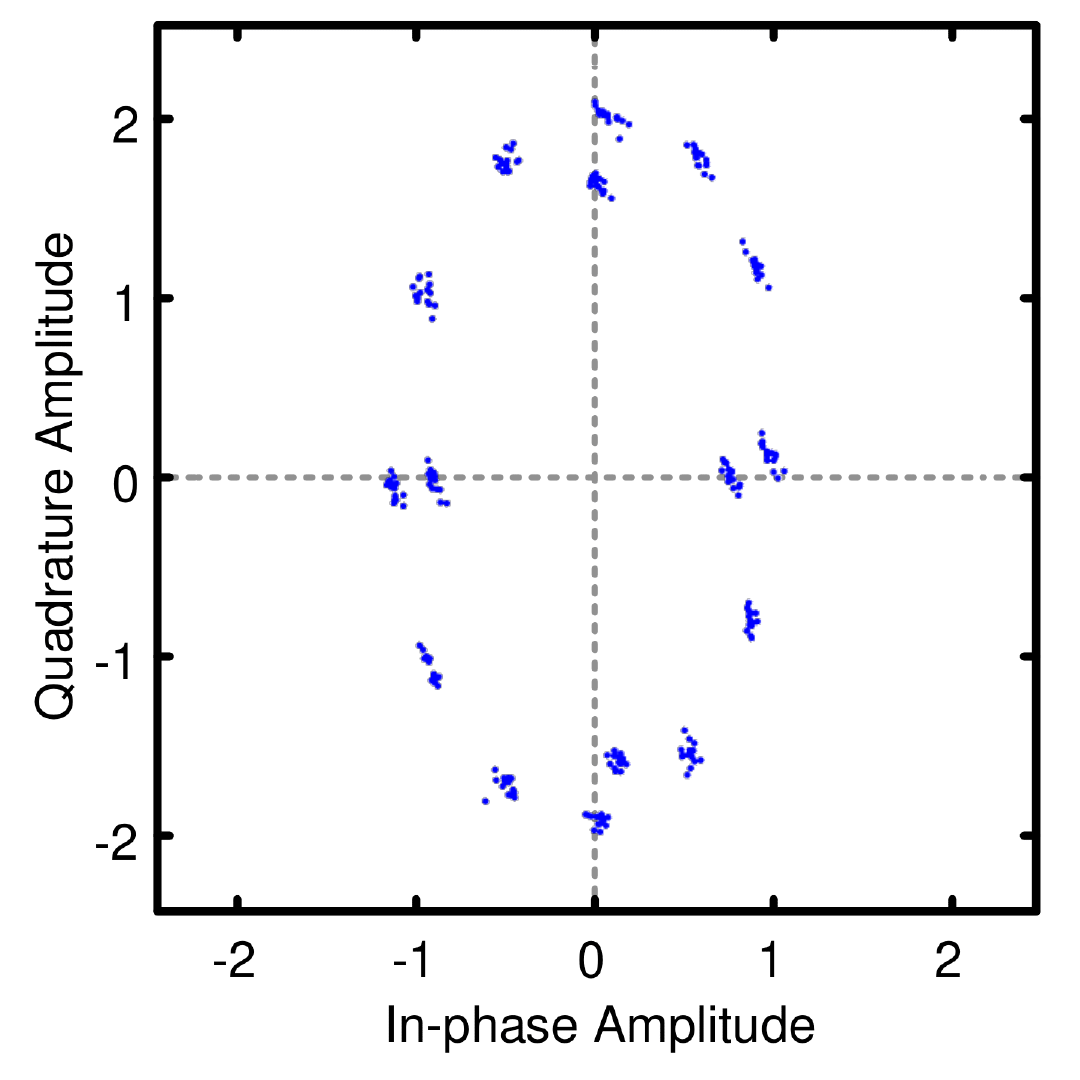} \label{16-PSK1}}

\caption{ 
Reconstructed constellations at the
receiver in the office space for four modulation schemes: 
 \protect\subref{BPSK1} BPSK \protect\subref{QPSK1} QPSK \protect\subref{8-PSK1} 8-PSK \protect\subref{16-PSK1} 16-PSK.} 
\label{Modulation_Room}
\end{figure}

Fig.~\ref{Modulation_Antenna_Room} and Fig.~\ref{Modulation_Room} show the reconstructed constellations at the receiver in the anechoic chamber and office space, respectively. As expected, there are less distortions in the constellations measured in the anechoic chamber.

From Fig. \ref{Modulation_Antenna_Room}, an I/Q imbalance is evident, particularly for 16-PSK modulation. As the modulation order rises, it becomes necessary to utilize the codes that have components in the second harmonic. However, utilizing codes with components in the second harmonic leads to a reduction in amplitude in the first harmonic. To overcome this issue, it is essential to filter the received signals in the vicinity of the first harmonic. The implementation of such filtering was unattainable due to the narrow spectral gap between the first and second harmonics, attributed to the constraints of microcontroller and low switching speed. It is noteworthy that there are solutions to I/Q imbalance problem through further base band processing.

\subsection{Concurrent Data Modulation and Beam Steering}
We have demonstrated the feasibility of beam steering through bit shifting and data modulation via choosing proper codes from amplitude-phase pattern. In this part, we extend our exploration to showcase the capability of digital modulation concurrent with beam steering through this methodology. To achieve this, we developed a program to generate codes for QPSK modulation and subsequently applied them to the RIS cells.

\begin{figure}[t]
\centering
\includegraphics[width=3.5in]{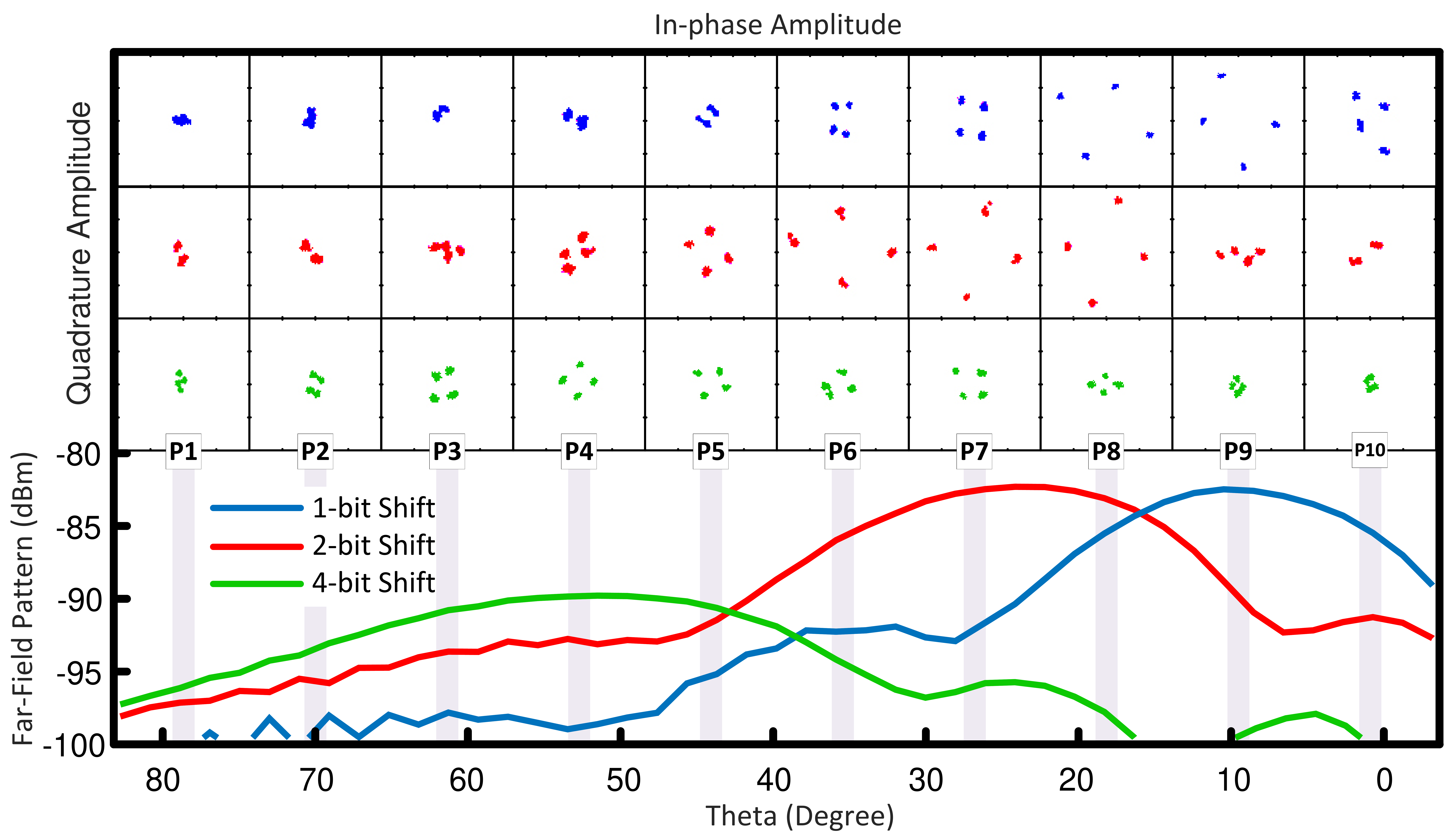}
\caption{Measured reflected pattern and constellation with respect to angular position for 1-bit, 2-bit, and 4-bit shifts into the adjacent cells of RIS columns. The constellations are measured at ten discrete angular positions identified as P1 to P10.
}
\label{const_steer}
\end{figure}

Fig.~\ref{const_steer} shows the measured reflected pattern and constellation with respect to angular
position for 1-bit, 2-bit, and 4-bit shifts into the adjacent cells of RIS columns all for the first harmonic. The QPSK constellations are measured at ten discrete angular positions. For this measurement, the RIS and Tx antenna are positioned on the rotary table. A distant radio receiver stores received signal at each angular rotation of the table. Upon completing the rotation, all recorded signals undergo processing to extract the received power in the first harmonic and the constellation diagrams at each angle. The degradation in the constellation diagram becomes increasingly pronounced with angular distance from the main beam as evident in Fig. \ref{const_steer}.

\section{Conclusion}
Our experimental results on a space-time coded RIS vividly illustrate the transmission of symbols during beam steering, and provide substantial evidence supporting the great potential for practical implementation of our proposed method in future wireless networks.
Development of reconfigurable intelligent surfaces with fast switching capabilities and parallel coding of cells are necessary steps toward practical deployment.

\bibliographystyle {IEEEtran}
\bibliography{refs}

\begin{thebibliography}{10}
\providecommand{\url}[1]{#1}
\csname url@samestyle\endcsname
\providecommand{\newblock}{\relax}
\providecommand{\bibinfo}[2]{#2}
\providecommand{\BIBentrySTDinterwordspacing}{\spaceskip=0pt\relax}
\providecommand{\BIBentryALTinterwordstretchfactor}{4}
\providecommand{\BIBentryALTinterwordspacing}{\spaceskip=\fontdimen2\font plus
\BIBentryALTinterwordstretchfactor\fontdimen3\font minus \fontdimen4\font\relax}
\providecommand{\BIBforeignlanguage}[2]{{%
\expandafter\ifx\csname l@#1\endcsname\relax
\typeout{** WARNING: IEEEtran.bst: No hyphenation pattern has been}%
\typeout{** loaded for the language `#1'. Using the pattern for}%
\typeout{** the default language instead.}%
\else
\language=\csname l@#1\endcsname
\fi
#2}}
\providecommand{\BIBdecl}{\relax}
\BIBdecl

\bibitem{agiwal2016next}
M.~Agiwal, A.~Roy, and N.~Saxena, ``Next generation {5G} wireless networks: A comprehensive survey,'' \emph{IEEE Communications Surveys \& Tutorials}, vol.~18, no.~3, pp. 1617--1655, 2016.

\bibitem{dehos2014millimeter}
C.~Dehos, J.~L. Gonz{\'a}lez, A.~De~Domenico, D.~Ktenas, and L.~Dussopt, ``Millimeter-wave access and backhauling: The solution to the exponential data traffic increase in {5G} mobile communications systems?'' \emph{IEEE Communications Magazine}, vol.~52, no.~9, pp. 88--95, 2014.

\bibitem{tariq2020speculative}
F.~Tariq, M.~R. Khandaker, K.-K. Wong, M.~A. Imran, M.~Bennis, and M.~Debbah, ``A speculative study on {6G},'' \emph{IEEE Wireless Communications}, vol.~27, no.~4, pp. 118--125, 2020.

\bibitem{harter2020generalized}
T.~Harter, C.~F{\"u}llner, J.~N. Kemal, S.~Ummethala, J.~L. Steinmann, M.~Brosi, J.~L. Hesler, E.~Br{\"u}ndermann, A.-S. M{\"u}ller, W.~Freude \emph{et~al.}, ``Generalized kramers--kronig receiver for coherent terahertz communications,'' \emph{Nature Photonics}, vol.~14, no.~10, pp. 601--606, 2020.

\bibitem{hosseininejad2018mac}
S.~E. Hosseininejad, S.~Abadal, M.~Neshat, R.~Faraji-Dana, M.~C. Lemme, C.~Suessmeier, P.~H. Bol{\'\i}var, E.~Alarc{\'o}n, and A.~Cabellos-Aparicio, ``{MAC}-oriented programmable terahertz phy via graphene-based yagi-uda antennas,'' in \emph{2018 IEEE Wireless Communications and Networking Conference (WCNC)}.\hskip 1em plus 0.5em minus 0.4em\relax IEEE, 2018, pp. 1--6.

\bibitem{hosseininejad2018reconfigurable}
S.~E. Hosseininejad, M.~Neshat, R.~Faraji-Dana, M.~Lemme, P.~Haring~Bol{\'\i}var, A.~Cabellos-Aparicio, E.~Alarc{\'o}n, and S.~Abadal, ``Reconfigurable {THz} plasmonic antenna based on few-layer graphene with high radiation efficiency,'' \emph{Nanomaterials}, vol.~8, no.~8, p. 577, 2018.

\bibitem{han2018ultra}
C.~Han, J.~M. Jornet, and I.~Akyildiz, ``Ultra-massive {MIMO} channel modeling for graphene-enabled terahertz-band communications,'' in \emph{2018 IEEE 87th vehicular technology conference (VTC Spring)}.\hskip 1em plus 0.5em minus 0.4em\relax IEEE, 2018, pp. 1--5.

\bibitem{amiri2018extremely}
A.~Amiri, M.~Angjelichinoski, E.~De~Carvalho, and R.~W. Heath, ``Extremely large aperture massive {MIMO}: Low complexity receiver architectures,'' in \emph{2018 IEEE Globecom Workshops (GC Wkshps)}.\hskip 1em plus 0.5em minus 0.4em\relax IEEE, 2018, pp. 1--6.

\bibitem{li2024}
P.~Li, E.~Zahedi, Y.~Shi, Y.~Deng, and L.~Liu, ``Dynamically tuning and reconfiguring microwave bandpass filters using optical control of switching elements,'' \emph{Microwave and Optical Technology Letters}, vol.~66, no.~2, p. e34082, 2024.

\bibitem{patron2014}
D.~Patron, A.~S. Daryoush, and K.~R. Dandekar, ``Optical control of reconfigurable antennas and application to a novel pattern-reconfigurable planar design,'' \emph{Journal of Lightwave Technology}, vol.~32, no.~20, pp. 3394--3402, 2014.

\bibitem{liu2021reconfigurable}
Y.~Liu, X.~Liu, X.~Mu, T.~Hou, J.~Xu, M.~Di~Renzo, and N.~Al-Dhahir, ``Reconfigurable intelligent surfaces: Principles and opportunities,'' \emph{IEEE Communications Surveys \& Tutorials}, vol.~23, no.~3, pp. 1546--1577, 2021.

\bibitem{elmossallamy2020reconfigurable}
M.~A. ElMossallamy, H.~Zhang, L.~Song, K.~G. Seddik, Z.~Han, and G.~Y. Li, ``Reconfigurable intelligent surfaces for wireless communications: Principles, challenges, and opportunities,'' \emph{IEEE Transactions on Cognitive Communications and Networking}, vol.~6, no.~3, pp. 990--1002, 2020.

\bibitem{basar2019wireless}
E.~Basar, M.~Di~Renzo, J.~De~Rosny, M.~Debbah, M.-S. Alouini, and R.~Zhang, ``Wireless communications through reconfigurable intelligent surfaces,'' \emph{IEEE Access}, vol.~7, pp. 116\,753--116\,773, 2019.

\bibitem{wu2019towards}
Q.~Wu and R.~Zhang, ``Towards smart and reconfigurable environment: Intelligent reflecting surface aided wireless network,'' \emph{IEEE Communications Magazine}, vol.~58, no.~1, pp. 106--112, 2019.

\bibitem{alamzadeh2023}
I.~Alamzadeh and M.~F. Imani, ``Detecting angle of arrival on a hybrid ris using intensity-only data,'' \emph{IEEE Antennas and Wireless Propagation Letters}, vol.~22, no.~9, pp. 2325--2329, 2023.

\bibitem{el2023}
M.~El-Absi, A.~A. Abbas, D.~Tubail, F.~Ilgac, A.~Abuelhaija, Y.~Zantah, S.~Ikki, A.~Sezgin, and T.~Kaiser, ``Path loss modeling of {RFID} backscatter channels with reconfigurable intelligent surface: experimental validation,'' \emph{IEEE Access}, 2023.

\bibitem{2aydin2023IEEE}
M.~Heinrichs, A.~Sezgin, and R.~Kronberger, ``Open source reconfigurable intelligent surface for the frequency range of 5 {GHz} {WiFi},'' in \emph{2023 IEEE International Symposium On Antennas And Propagation (ISAP)}.\hskip 1em plus 0.5em minus 0.4em\relax IEEE, 2023, pp. 1--2.

\bibitem{alamzadeh2022}
I.~Alamzadeh and M.~F. Imani, ``Sensing and reconfigurable reflection of electromagnetic waves from a metasurface with sparse sensing elements,'' \emph{IEEE Access}, vol.~10, pp. 105\,954--105\,965, 2022.

\bibitem{aydin2023IEEE}
S.~Tewes, M.~Heinrichs, K.~Weinberger, R.~Kronberger, and A.~Sezgin, ``A comprehensive dataset of {RIS}-based channel measurements in the 5{GHz} band,'' in \emph{2023 IEEE 97th Vehicular Technology Conference (VTC2023-Spring)}, 2023, pp. 1--5.

\bibitem{yang2022terahertz}
F.~Yang, P.~Pitchappa, and N.~Wang, ``Terahertz reconfigurable intelligent surfaces ({RISs}) for {6G} communication links,'' \emph{Micromachines}, vol.~13, no.~2, p. 285, 2022.

\bibitem{aboagye2022ris}
S.~Aboagye, A.~R. Ndjiongue, T.~M. Ngatched, O.~A. Dobre, and H.~V. Poor, ``{RIS}-assisted visible light communication systems: A tutorial,'' \emph{IEEE Communications Surveys \& Tutorials}, vol.~25, no.~1, pp. 251--288, 2022.

\bibitem{cui2014coding}
T.~J. Cui, M.~Q. Qi, X.~Wan, J.~Zhao, and Q.~Cheng, ``Coding metamaterials, digital metamaterials and programmable metamaterials,'' \emph{Light: Science \& Applications}, vol.~3, no.~10, pp. e218--e218, 2014.

\bibitem{zhang2018space}
L.~Zhang, X.~Q. Chen, S.~Liu, Q.~Zhang, J.~Zhao, J.~Y. Dai, G.~D. Bai, X.~Wan, Q.~Cheng, G.~Castaldi \emph{et~al.}, ``Space-time-coding digital metasurfaces,'' \emph{Nature Communications}, vol.~9, no.~1, p. 4334, 2018.

\bibitem{yang2020coverage}
L.~Yang, Y.~Yang, M.~O. Hasna, and M.-S. Alouini, ``Coverage, probability of {SNR} gain, and {DOR} analysis of {RIS}-aided communication systems,'' \emph{IEEE Wireless Communications Letters}, vol.~9, no.~8, pp. 1268--1272, 2020.

\bibitem{yu2019miso}
X.~Yu, D.~Xu, and R.~Schober, ``{MISO} wireless communication systems via intelligent reflecting surfaces,'' in \emph{2019 IEEE/CIC International Conference on Communications in China (ICCC)}.\hskip 1em plus 0.5em minus 0.4em\relax IEEE, 2019, pp. 735--740.

\bibitem{di2020hybrid}
B.~Di, H.~Zhang, L.~Song, Y.~Li, Z.~Han, and H.~V. Poor, ``Hybrid beamforming for reconfigurable intelligent surface based multi-user communications: Achievable rates with limited discrete phase shifts,'' \emph{IEEE Journal on Selected Areas in Communications}, vol.~38, no.~8, pp. 1809--1822, 2020.

\bibitem{huang2019reconfigurable}
C.~Huang, A.~Zappone, G.~C. Alexandropoulos, M.~Debbah, and C.~Yuen, ``Reconfigurable intelligent surfaces for energy efficiency in wireless communication,'' \emph{IEEE Transactions on Wireless Communications}, vol.~18, no.~8, pp. 4157--4170, 2019.

\bibitem{hadad2016breaking}
Y.~Hadad, J.~C. Soric, and A.~Alu, ``Breaking temporal symmetries for emission and absorption,'' \emph{Proceedings of the National Academy of Sciences}, vol. 113, no.~13, pp. 3471--3475, 2016.

\bibitem{hadad2015space}
Y.~Hadad, D.~L. Sounas, and A.~Alu, ``Space-time gradient metasurfaces,'' \emph{Physical Review B}, vol.~92, no.~10, p. 100304, 2015.

\bibitem{dai2018independent}
J.~Y. Dai, J.~Zhao, Q.~Cheng, and T.~J. Cui, ``Independent control of harmonic amplitudes and phases via a time-domain digital coding metasurface,'' \emph{Light: Science \& Applications}, vol.~7, no.~1, p.~90, 2018.

\bibitem{Liu21}
Y.~Liu, L.~Wu, Z.~Zhang, J.~Dang, and B.~Zhu, ``Space-time coding design for {RIS}-assisted integrated sensing and communication system,'' \emph{IEEE Wireless Communications Letters}, vol.~13, no.~7, pp. 1943--1947, 2024.

\bibitem{Pang22}
Y.~Pang, X.~Lei, Y.~Xiao, and H.~Niu, ``Reconfigurable intelligent surface assisted space-time line code for {SIMO} transmission,'' \emph{IEEE Communications Letters}, vol.~26, no.~12, pp. 3069--3073, 2022.

\bibitem{gholami2022direct}
M.~Gholami and M.~Neshat, ``A direct antenna modulator with beam steering capability based on space-time-coding arrays,'' \emph{IEEE Transactions on Antennas and Propagation}, vol.~70, no.~10, pp. 9282--9291, 2022.

\bibitem{kaina}
N.~Kaina, M.~Dupr{\'e}, M.~Fink, and G.~Lerosey, ``Hybridized resonances to design tunable binary phase metasurface unit cells,'' \emph{Optics Express}, vol.~22, no.~16, pp. 18\,881--18\,888, 2014.

\end{thebibliography}

\end{document}